\documentclass[twocolumn,showpacs,preprintnumbers,amsmath,amssymb,prl]{revtex4}
\usepackage{graphicx,epsfig}% Include figure files
\usepackage{dcolumn}% Align table columns on decimal point
\usepackage{bm}% bold math
\def\alt{\mathrel{\hbox{\rlap{\hbox{\lower3pt\hbox{$\sim$}}}
\hbox{\raisebox{0.4ex}{\hspace*{-0.05in}$<$}}}}}
\def\agt{\mathrel{\hbox{\rlap{\hbox{\lower4pt\hbox{$\sim$}}}
\hbox{\raisebox{0.4ex}{\hspace*{-0.05in}$>$}}}}}

\begin{document}

\title{Neural decision boundaries for maximal information transmission}

\author{Tatyana Sharpee$^a$ and William Bialek$^b$} 
%\email{sharpee@phy.ucsf.edu;wbialek@princeton.edu }
\affiliation{ $^a$Sloan--Swartz Center for Theoretical Neurobiology and
Department of Physiology, University of California at San Francisco,
San Francisco, CA 94143\\
$^b$Joseph Henry Laboratories of Physics, 
Lewis--Sigler Institute for Integrative Genomics, and the Princeton Center for Theoretical Physics,
 Princeton
University, Princeton, New Jersey 08544}

\date{\today}
\begin{abstract}
  We consider here how to separate multidimensional signals
  into two categories, such that the binary decision transmits the maximum possible information
  transmitted about those signals. Our motivation comes from the
  nervous system, where neurons process multidimensional signals into
  a binary sequence of responses (spikes). In a small noise limit, we 
  derive a general equation for the
  decision boundary that locally relates its curvature to the probability distribution of inputs. We show
  that for Gaussian inputs the optimal boundaries are planar, but for non--Gaussian inputs the curvature is nonzero.  As an example, we consider exponentially
  distributed inputs, which are known to approximate a variety of
  signals from natural environment.  
\end{abstract}

\maketitle

What we know about the world around us is represented in the nervous system by sequences of discrete electrical pulses termed action potentials or ``spikes'' \cite{spikes}.  One attractive theoretical idea, going back to the 1950s, is that these representations constructed by the brain are efficient in the sense of information theory \cite{attneave_54,barlow_59,barlow_61}.  These ideas have been formalized to predict the spatial and temporal filtering properties of neurons \cite{srinivasan+al_82,buchsbaum+gottschalk_83,field_87,linsker_89,atick+redlich_90}, as well as the shapes of nonlinear input/output relations \cite{laughlin_81}, showing how these measured behaviors of cells can be understood as optimally matched to the statistical properties of natural sensory inputs.  There have been attempts, particularly in the auditory system, to test directly the prediction that the coding of naturalistic inputs is more efficient \cite{rieke+al_95,nelken+al_99,escabi+al_03,machens+al_03,hsu+al_04}, and this concept of matching has been used also to predict new forms of adaptation to the input statistics \cite{smirnakis+al_97,brenner+al_00,fairhall+al_01,david+al_04,sharpee+al_06,maravall+al_07}.    
Despite this progress, relatively little attention has been given to the problem of optimal coding in the presence of the strong, threshold--like nonlinearities associated with the generation of spikes \cite{deweese_96}.

Sensory inputs to the brain are intrinsically high dimensional objects.
For example, visual neurons encode various patterns of light intensities that, upon
moderate discretization, become vectors in $10^2 - 10^3$ dimensional
space \cite{sharpee+al_04,bialek+ruyter_05}. We can think of the ``decision'' to generate an action potential as drawing boundaries in  these high dimensional spaces, so that a theory of optimal coding for spiking neurons is really a theory for the shape of these boundaries.
In the simplest perceptron--like models \cite{minsky+papert_69}, boundaries are planar, and spiking thus is determined by only a single (Euclidean) projection of the stimulus onto a vector normal to the dividing plane.  In the perceptron limit, the optimal choice of decision boundaries reduces to the choice of an optimal linear filter.  But a number of recent experiments suggest that neurons, even in early stages of sensory processing, are sensitive to multiple stimulus projections, with intrinsically curved decision boundaries \cite{brenner+al_00,bialek+ruyter_05,arcas+al_03,rust+al_05, slee+al_05,felsen+al_05,fairhall+al_06}.  Here we try to develop a theory of optimal coding for spiking neurons in which these curved boundaries emerge naturally.

We consider a much simplified version of the full problem.  We look at a single neuron, and focus on a small window of time in which that cell either does or does not generate an action potential.  We ignore, in this first attempt, coding strategies that involve patterns of spikes across multiple neurons or across time in single neurons, and ask simply how much information the binary spike/no spike decision conveys about the input signal.  Let this input signal be a vector $\bf r$ in a space of $d$ dimensions \cite{history} and let the distribution of these signals be given by $P({\bf r})$.  If the binary output of the neuron is $\sigma$, we are interested in calculating the mutual information $I(\sigma ; {\bf r})$ between $\sigma$ and the input $\bf r$.

We can write the information as a difference between two entropies
\cite{spikes,cover+thomas}, the response entropy and the noise
entropy:
$I(\sigma ; {\bf r}) = H_{\rm response}-H_{\rm noise}$.
In our simplified problem, with a single neuron giving binary responses, the response entropy,
\begin{equation}
H_{\rm response}=-p\log p-(1-p) \log (1-p) ,
\end{equation}
 is completely determined by
the average spike probability $p$.  We might imagine that this  probability  is set by constraints outside the problem of coding itself.  For example, generating spikes costs energy, and so metabolic constraints might fix the mean spike rate \cite{laughlin+al_98,sarpeshkar_98,atwell+laughlin_01,balasubramanian+al_01}.  Our problem, then, is to find coding strategies that minimize the noise entropy at fixed $p$.

In the absence of noise, the coding scheme which maps signals into spikes (or not) is a boundary in the $d$--dimensional space of inputs.  If the domain in which spikes occur is $G$, then   $p=\int_G d^d {\bf r} P({\bf r})$.  If the noise truly were zero, all codes with the same value of $p$ would transmit the same amount of information, and there would be an infinite set of nominally optimal domains $G$.

We will work in an approximation where the noise is small and additive with the input.  Then if the boundary of the spiking domain is some ($d-1$ dimensional) surface $\gamma$, we expect that responses far from this boundary are essentially deterministic and do not contribute to the noise entropy; all of the contributions to $H_{\rm noise}$ should arise from a narrow strip surrounding the boundary $\gamma$.  Within this strip, the response is almost completely uncertain.  Thus we can approximate the noise entropy by saying that is $\sim 1\,{\rm bit}$ inside the strip, and zero outside; the total noise entropy is then the mass of probability inside the strip.  The width of the strip is proportional to the strength of the noise, and if noise is small the probability distribution of inputs does not vary significantly across this width, so we can write the overall noise entropy as an integral along the
decision boundary $\gamma$:
\begin{equation}
\label{noise-entropy}
H_{\rm noise}\left[\gamma; P({\bf r})\right]\approx\sigma \int_\gamma ds  P({\bf r}),
\end{equation}
where $ds$ is the infinitesimal surface element of dimension $d-1$ on
the decision boundary $\gamma$ and $\sigma$ is the amplitude of the noise. The exact shape of the nonlinear
function describing how spike probability changes across the domain
boundary might introduce additional numerical factor of order unity in
Eq~(\ref{noise-entropy}), but these can always be incorporated into
defining $\sigma$ as the {\em effective}
noise level.

While our choice of threshold--like transitions between spiking and
non--spiking regions considerably narrows the types of possible
input-output transformations, it still leads, as we show below, to
highly nontrivial, yet tractable, solutions. We will treat the noise
length scale $\sigma$ as one of the pre-defined parameters; it can
take arbitrary positive values and will set the units for measuring
contours' curvature and $\nabla \ln P({\bf r})$.

Taking into account that the response entropy $H_{\rm response}$ only
depends on the average spike probability, the optimal contour
providing maximal information may be found by minimizing
\begin{equation}
\label{functional}
F= \sigma \int_\gamma ds  P({\bf r})-\lambda\left [p-\int_G d^dr P({\bf r})\right],
\end{equation}
where $\lambda$ is the Lagrange multiplier incorporating the
constraint for the average spike probability $p$. For an optimal
contour, first order variation of $F$ respect to local perturbations
$\delta {\bf r}$ in the contour's shape should be zero.  Only
perturbations along the surface normal could  change the
value of the functional. 
Both  the infinitesimal  surface  element and  the probability  values
along the decision boundary are subject to change:
\begin{equation}
\delta F = \int_\gamma ds\delta r_\perp(s)\left[ 
\sigma   {\bf \hat t_{\rm i}}\cdot \frac{d{\bf \hat n}}{ds_{\rm i}} P({\bf r}) +\sigma {\bf \hat n}\cdot  \nabla P+\lambda  P({\bf r})\right],
\end{equation}
where the set of vectors $ \{ {\bf \hat t}_1 ,\, {\bf \hat t}_2, \cdots , \, {\bf \hat t}_{d-1} \}$   defines the tangent plane, ${\bf \hat n}$ defines the normal, and we use the summation convention for the index $\rm i$.  Because perturbations at various points along the surface are
independent, the optimal contour should satisfy: 
\begin{eqnarray}
\label{local}
\lambda+ \kappa + {\bf \hat n}\cdot \nabla \ln P &=&0, \\
\quad \kappa={\bf \hat t}_{\rm i}\cdot d{\bf \hat n}/{ds_{\rm i}}&=&{\bf div}\, {\bf \hat n},
\end{eqnarray}
where $\kappa$ is the mean curvature of the decision boundary
$\gamma$.  Notice that we have rescaled $\lambda$ by a factor of $\sigma$, so there is only one parameter in the problem. 

Below we solve Eq~(\ref{local}) to find optimal decision boundaries
for two example probability distributions: a Gaussian and the
exponential.  The exponential distribution is important not only as an
example of non--Gaussian inputs, but also because it captures some of
the essential statistical properties found in real--world signals
\cite{ruderman+bialek_94,ruderman_94}.

\begin{figure}[t]
\includegraphics[width=3.1in]{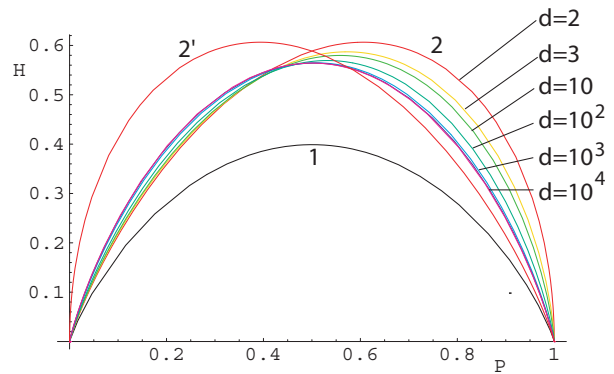}
\caption{Comparison of noise entropies for
  straight line solutions (1) and circles with spiking on the outside
  (2) or inside (2') for Gaussian inputs. The entropy for circular
  solution depends on the dimensionality $d$ of inputs as illustrated
  here in the case of spiking outside of a circle.\label{fig:HP_gaussian} }
\end{figure}

Consider the case of
uncorrelated Gaussian inputs 
$P({\bf r})=(2\pi)^{-d/2 } \exp(-r^2/2)$, where the equation on
optimal contours takes the form:
\begin{equation}
\label{local_gauss}
\lambda +\kappa -{\bf \hat n}\cdot {\bf r}=0.
\end{equation}
The families of possible solution include circles
[$\lambda=r-(d-1)/r$, where $r$ is the circle radius] and straight
lines $\kappa=0$ [$\lambda=r$, where $r$ is the smallest distance from
the line to the origin]. 
Circles and straight lines turn out to be the only possible smooth contours \cite{contours}.

To choose between circles and straight lines, we calculate
the noise entropy as a function of spike probability $p$ in both cases. From Eq (\ref{noise-entropy}), we see that $H_{\rm noise}$ is proportional to the noise level $\sigma$, so in what follows we compute the noise entropy in these units.
For straight lines a distance $r$ from the origin, $H_{\rm line}=
\exp(-r^2/2)/\sqrt{2\pi}$ and $P_{\rm line}=\left[1-{\rm
    erf}(r/\sqrt{2})\right]/2$. For a circle in two dimensions
$H_{\rm circle}=r\exp(-r^2/2)$ and $P_{\rm outside-a-
  circle}=\exp(-r^2/2)$.  Expressions for entropy and probability for
straight lines do not change with dimensionality $d$, while the
corresponding values for circles are: 
\begin{eqnarray}
H_{\rm circle}^{(d)} &=&
\frac{2r^{d-1}e^{-r^2/2}}{2^{d/2}\Gamma(\frac{d}{2})},\\
 P_{\rm
  outside-a-
  circle}^{(d)}&=&\frac{\Gamma(\frac{d}{2},\frac{r^2}{2})}{\Gamma(\frac{d}{2})},
  \end{eqnarray}
where $\Gamma(n,x)=\int_x^{\infty} dt e^{-t} t^{n-1}$ is the
incomplete Gamma function. In Fig.~\ref{fig:HP_gaussian} we plot these
solutions to show that for any probability $p$ and dimensionality $d$,
the optimal separation  is with straight boundaries.
This result also holds for correlated Gaussian inputs, where the optimal hyperplane is
the one which intersects the axis of largest variance and is parallel
to other coordinate axes.

As an example of a non--Gaussian
probability distribution, we consider an exponential distribution in
two dimensions (2D): $P(x,y)=\frac{1}{4}e^{-|x|-|y|}$.  The local
equation for optimal contours (\ref{local}) can be written
parametrically:
\begin{eqnarray}
\label{expo_local}
&&\frac {d\phi}{ds}=\lambda +\sin \phi- \cos \phi, \quad x>0, y>0\\
&&\frac {dx}{ds}=\cos\phi, \quad \frac {dy}{ds}=\sin\phi \nonumber
\end{eqnarray}
where angle $\phi$ determines the tangent ${\bf \hat t}=(\cos\phi,
\sin\phi)$ and normal ${\bf \hat n}=(-\sin \phi ,\cos\phi)$ of the curve, as
well as the curvature $\kappa=-{d\phi}/{ds}$. Solutions in other
quadrants can be obtained from Eq (\ref{expo_local}) by an appropriate
change of variables.

\begin{figure}[t]
\includegraphics[width=3.3in]{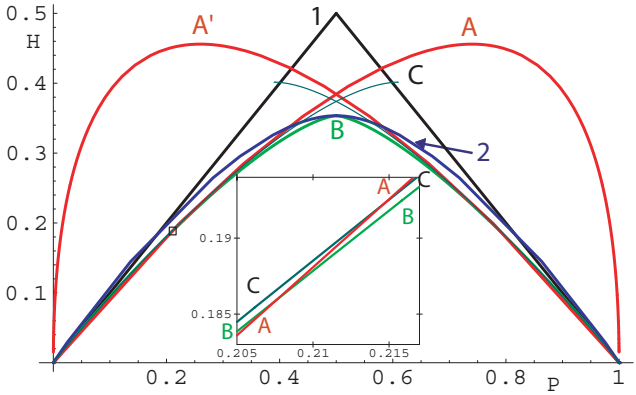}
\caption{Noise entropy
  $H$ along various decision boundaries with exponential inputs:
  ``streched circles'' with spiking outside A or inside A', extended
  solutions B and C; straight lines parallel to coordinate axes (1)
  and at $\pm \pi/4$ angle (2).  Solutions (A-C) and (1) satisfy the
  optimality Eq~(\protect\ref{expo_local}), but not (2), which is
  optimal only at a single point at $p=1/2$ where it becomes part of
  family of extended solutions B.  Inset shows that switching occurs
  between solutions A and B.\label{fig:entropy_probability_expo}}
\end{figure}

For $\lambda=\pm 1$, the family of optimal contours includes straight
lines parallel to coordinate axes. Such straight lines represent 1D
threshold decisions, and in this case the noise entropy equals the
spike probability, decreasing exponentially with the threshold $r$ for
decision $x>r$:
\begin{equation}
\label{1Dthreshold}
H_{\rm independent}= P_{\rm independent}=e^{-r}/2.
\end{equation}
The only other straight line solution that satisfies the optimality
condition in Eq (\ref{expo_local}) is a line $y=\pm x$; it corresponds to
spike probability $p=1/2$. Straight lines of the same angle that do
not pass through the origin do not satisfy the optimality condition,
but they provide a useful benchmark for other solutions in the middle
range of probabilities $0.2 \alt p\alt 0.8$, where they are better
than the straight lines parallel to the axes:
$H_{\pi/4}=\sqrt{2}(r+1)\exp(-r)/4$,
$P_{\pi/4}=(r+2)\exp(-r)/4$, as Fig.~\ref{fig:entropy_probability_expo}
illustrates.

\begin{figure}[t]
\includegraphics[width=3.1in]{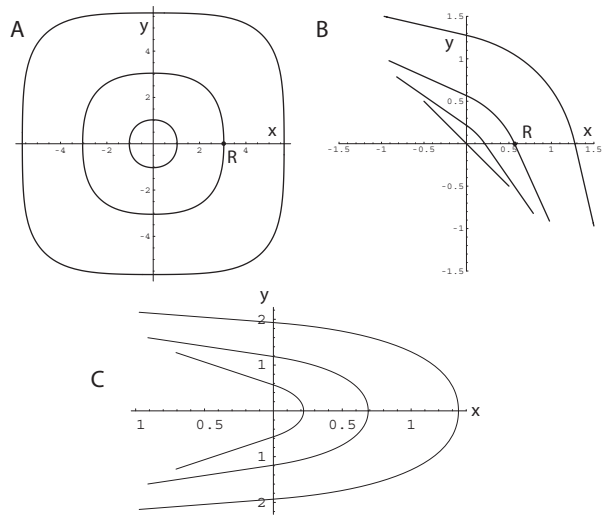}
\caption{\label{fig:family_expo} Optimal solutions for 2D exponential
  inputs: (A) closed ``stretched circle'' solutions are shown for
  $\lambda=-0.3,-0.9,-0.99$, numbers correspond to the increasing size
  of the curved segment throughout this legend. (B) Extended solutions
  symmetric around $y=x$ line are shown for
  $\lambda=0,\,-0.25,\,-0.5,\,-0.75$. (C) Extended solutions symmetric
  around $x=0$ line are shown for $\lambda=0,\, -0.5,\,-0.75$ [this
  type turned out to be suboptimal, albeit by a small margin, compared
  to either A or B, cf. Fig.~\ref{fig:entropy_probability_expo}].}
\end{figure}

Within a single quadrant, the optimal solution can be found explicitly in
terms of angle $\phi$ relative to the starting point where
$\phi=\phi_0$, $x_0=x(\phi_0)$, and $y_0=y(\phi_0)$:
\begin{eqnarray}
\label{within_quadrant}
\hspace*{-0.35in}&&x(\phi)+y(\phi)=x_0+y_0+\ln\frac{\lambda +\sin \phi -\cos \phi}{\lambda +\sin \phi_0 -\cos \phi_0}\nonumber \\
\hspace*{-0.35in}&&y(\phi)-x(\phi)=y_0-x_0+ \phi-\phi_0-\lambda\left[ s(\phi)-s(\phi_0)\right],
\end{eqnarray}
where arc length $s(\phi)$ depends on the angle $\phi$ as:

\begin{eqnarray}
\label{arc}
&& s(\phi)= \left \{\begin{array}{ll} 
\frac{1}{\sqrt{2-\lambda^2}}\ln \frac{u(\phi) -1}{u(\phi)+1}, & |\lambda|<\sqrt{2} \\ 
\frac{2}{\sqrt{\lambda^2-2}}\tan^{-1}u(\phi) , 
& |\lambda|>\sqrt{2} 
\end{array} \right.,  \\
&& u(\phi)=\left(1+(\lambda+1)\tan (\phi/2)\right)/\sqrt{|\lambda^2-2|}. \nonumber
\end{eqnarray}
These solutions are similar to a logarithmic spiral for
$|\lambda|>\sqrt{2}$, and to a hyperbola for $|\lambda|<\sqrt{2}$,
with asymptotes at $\pi/4-\arcsin (\lambda /\sqrt{2})$ and
$5\pi/4+\arcsin(\lambda /\sqrt{2})$.  Asymptotes themselves are
valid solutions within a quadrant; they will be part of a global
solution.  For all $\lambda$ the solution (\ref{within_quadrant})
intersects coordinate axes where it should be matched with similar
solutions in other quadrants.  

The possible types of global solutions are shown in
Fig.~\ref{fig:family_expo}. They could be either closed (``stretched circles''; A) or
extended (B and C) \cite{nocurves}. For $|\lambda|<\sqrt{2}$, extended solutions
can be formed by connecting asymptotes  
in two separate quadrants with a convex curve described by
Eq's~(\ref{within_quadrant},\ref{arc}). We will refer to such extended
solutions as B or C depending upon whether the curved segment passes
through one or two quadrants, cf. Fig.~\ref{fig:family_expo}.
Extended solutions B are symmetric around $y=x$ line, and exist only
for $-1<\lambda<0$, while
extended solutions C are
symmetric around $x=0$ line, and exist for  $-1< \lambda <1$.

For all types of global solutions (A-C), boundary conditions specify a
unique curve for each value of $\lambda$. In all cases, both entropy
and probability can be found exactly as a function of $\lambda$. For
solutions A we find  
\begin{eqnarray}
H_A&=&\frac{2-\lambda(\lambda+1)\Delta s_A}{2-\lambda^2} \, e^{-R_A},\\
P_A
&=&\left[1+\frac{(\lambda+1)\Delta
    s_A-\lambda}{2-\lambda^2}\right]e^{-R_A},
    \end{eqnarray}
with expressions for
arc length $\Delta s_A$ and size of the curved segment $R_A$ from
\cite{expressions}.
For extended solutions B, the entropy and probability become   \cite{extended}
\begin{eqnarray}
H_B&=&e^{-R_B}\frac{
  2(\sqrt{2-\lambda^2}-\lambda) - \lambda \Delta
  s_B(\lambda+\sqrt{2-\lambda^2})}{4(2-\lambda^2)}\nonumber\\
  &&
\\
P_B&=&\frac{4-\lambda^2-\lambda\sqrt{2-\lambda^2}+\Delta
  s_B(\lambda+\sqrt{2-\lambda^2})}{4(2-\lambda^2)} \, e^{{-R_B}} .\nonumber\\
  &&
\end{eqnarray}
More detailed calculations shows that solutions C are suboptimal compared to
global solutions A or B;  see
Fig.~\ref{fig:entropy_probability_expo} and the  discussion below. Note
that neither A, B, nor C solutions exist for $\lambda<-1$.

The most physiologically relevant regime corresponds to $\lambda
=-1+\epsilon$, $\epsilon\ll 1$. Here, all global solutions A-C have a
large ``radius''of the curved segment $R\sim-\ln\epsilon$. The
probability and noise entropy depend exponentially on $R$, so that
$P \sim \sqrt{2}e^{-\pi/4}\left( \epsilon-3\epsilon^2\ln
\epsilon\right) +\alpha\epsilon^2 $ and
$H/P \sim 1-\frac{\epsilon}{2}
+\frac{\epsilon^2}{2}\ln \epsilon +\beta \epsilon^2$. The
constants $\alpha$ and $\beta$ depend on the solution type
(A-C). Because $\beta_A<\beta_C<\beta_B$, solutions $A$ are optimal
for small $\epsilon$.  Near $p \approx 0.2$, intersections between the
three curves occur.  In the $O(\epsilon^2)$ approximation, all of the
three curves intersect at a single intersection point that
splits into three once higher-order terms are included.  As probability
increases, B and C intersect first (A goes below), then A and B
(the crossover point, C goes above), and finally, A and C (B
goes below). The inset of Fig.~\ref{fig:entropy_probability_expo}
shows A-B and A-C intersections. Thus, solutions A and B are optimal
at extreme and medium probabilities, respectively. Solutions of type C
are never optimal, and neither are the straight line solutions, except
for the middle point $p=1/2$.

In summary, we have presented a general approach to finding
optimal binary separations of multidimensional inputs. In the small noise limit, the curvature
of the optimal bounding surface is determined locally by the probability
distribution. While Gaussian inputs are optimally separated by hyperplanes, this is not the case in general. For example, in the
case of exponentially distributed inputs in two dimensions, the optimal decision
contours are curved and could either be
closed or extended.  Closed contours are optimal at extreme
probabilities, while extended ones are optimal for spike probabilities
near $1/2$.  The ubiquity of non--Gaussian signals in nature,
particularly of the exponential distributions considered here, suggests that
these results will be relevant for neurons across different sensory
modalities.

\begin{acknowledgments}
This work was supported in part by the NIMH and by the Swartz Foundation.
\end{acknowledgments}


\begin{thebibliography}{99}
\bibitem{history}
Note that what we call ``the input signal'' could in fact reflect the recent history of the physical inputs; we are interested in all aspects of the input that are (potentially) relevant to the question of whether a single neuron will generate an action potential in some small window of time.
\bibitem{contours}
Other smooth contours are not possible because
the mean curvature
increases exponentially with distance ($\kappa  \propto e^{r^2/2}$), causing contours to
self--intersect before a closed smooth contour can be obtained.
\bibitem{nocurves}
For the 2D exponential
  distribution, no curved solutions that extend to infinity and are
  confined to one or two quadrants may be better than the $H=P$
  solution achievable by straight lines parallel to the axes
  (\ref{1Dthreshold}).  This is due to the arc length 
  $\sqrt{1+(y'(x))^2}$ in the noise entropy 
  \begin{eqnarray}
  H&=&\int_r^\infty
  \hspace*{-0.1in}dx \sqrt{1+{y'}^2}e^{-x-y(x)},\\
  P&=&\int_r^\infty\hspace*{-0.1in} dx\, e^{-x-y(x)},
  \end{eqnarray}
   which makes $H>P$
  for all such solutions. This argument does not apply to solution
  spanning three quadrants or four quadrants, shown
  Fig.~\ref{fig:family_expo}.
\bibitem{expressions}
For solutions of type A, the arc length 
\begin{equation}
\Delta
  s_A=\frac{1}{\sqrt{2-\lambda^2}}\ln
  \frac{\lambda+2+\sqrt{2-\lambda^2}} {\lambda+2-\sqrt{2-\lambda^2}},
\end{equation}
  for $-1<\lambda \leq \sqrt{2}$, and 
\begin{equation}
  \Delta
  s_A=\frac{2}{\sqrt{\lambda^2-2}}\left(\frac{\pi}{2}-\tan^{-1}\frac{\lambda+2}{\sqrt{\lambda^2-2}}\right),
\end{equation}
  for $\lambda \geq \sqrt{2}$. The  size of the curved
  segment is $R_A=\pi/4-\lambda \Delta s_A/2$. The solutions are
  continuous at $\lambda=\sqrt{2}$.
\bibitem{extended}
For extended solutions of type B, the arc length 
\begin{equation}
\Delta
  s_B=-\frac{\ln (1+\lambda \sqrt{2-\lambda^2})}{\sqrt{2-\lambda^2}} ;
\end{equation}
  solutions are valid only for $-1<\lambda<0$. The corresponding size of the
  curved segment is $R_B=-\arcsin(\lambda/\sqrt{2})-\lambda \Delta
  s_B/2$.
\end{thebibliography}
\end{document}